\documentclass[journal]{IEEEtran}
\usepackage{float}
\usepackage{bbm}
\usepackage{footmisc}

\usepackage{multirow}
\usepackage{amssymb}
\usepackage{amsmath}
\usepackage{mathrsfs}
\usepackage{amsfonts}
\usepackage{graphicx}
\usepackage{cite}
\usepackage{subfigure}
\usepackage{tabularx}
\usepackage{epstopdf}
\usepackage{microtype}
\usepackage{stfloats}

\usepackage[noend]{algpseudocode}
\usepackage{algorithmicx,algorithm}
\usepackage{color}
\newtheorem{theorem}{Theorem}

\newtheorem{remark}{Remark}

\allowdisplaybreaks[4]

\makeatletter

\newcommand{\Rmnum}[1]{\expandafter\@slowromancap\romannumeral #1@}

\IEEEoverridecommandlockouts

\begin{document}

\title{\LARGE{Secrecy Offloading Rate Maximization for Multi-Access Mobile Edge Computing Networks}
\author{Mingxiong Zhao, Huiqi Bao, Li Yin, Jianping Yao, and Tony Q. S. Quek}
\thanks{M. Zhao, H. Bao, and L. Yin are with Engineering Research Center of Cyberspace, National Pilot School of Software, Yunnan University, Kunming, China. E-mail: mx\_zhao@ynu.edu.cn.}
\thanks {J. Yao is with the School of Information Engineering, Guangdong University of Technology, Guangzhou, China. E-mail: yaojp@gdut.edu.cn.}
\thanks{T. Q. S. Quek is with Singapore University of Technology and Design, Singapore. E-mail: tonyquek@sutd.edu.sg.}
}

\maketitle

\begin{abstract}
This letter considers a multi-access mobile edge computing (MEC) network consisting of multiple users, multiple base stations, and a malicious eavesdropper. Specifically, the users adopt the partial offloading strategy by partitioning the computation task into several parts. One is executed locally and the others are securely offloaded to multiple MEC servers integrated into the base stations by leveraging the physical layer security to combat the eavesdropping. We jointly optimize power allocation, task partition, subcarrier allocation, and computation resource to maximize the secrecy offloading rate of the users, subject to communication and computation resource constraints. Numerical results demonstrate that our proposed scheme can respectively improve the secrecy offloading rate 1.11\%--1.39\% and 15.05\%--17.35\% (versus the increase of tasks' latency requirements), and 1.30\%--1.75\% and 6.08\%--9.22\% (versus the increase of the maximum transmit power) compared with the two benchmarks. Moreover, it further emphasizes the necessity of conducting computation offloading over multiple MEC servers.
\end{abstract}


\vspace{-0.3cm}
\section{Introduction}
With the continuous promotion of fifth generation network, it has changed our life dramatically and spawned many new applications, such as autonomous driving, augmented reality and virtual reality, which need much higher quality-of-service (QoS) requirements than the traditional applications.
However, it is quite a challenge for mobile users with low computation capability and limited battery life to meet the stringent QoS requirements.
Hence, mobile edge computing (MEC) has been proposed and regarded as an enabling technology to overcome these challenges, by allowing mobile users to wirelessly offload their computation tasks to nearby base stations (BSs) equipped with MEC servers.

Due to the random nature of the wireless channel, mobile users can experience unfavorable instantaneous channel conditions to the BS. To tackle this, multi-access technology has emerged as a promising paradigm to leverage the diversity order to enhance the access capacity. In addition, it can enable mobile users to efficiently exploit the rich computation resource at the BS for reducing the local energy consumption and the processing latency (see \cite{Guo2018Computation,Zhao2021TVT,Song2021Joint}).

Moreover, due to the open nature of the wireless channel, communication security is another major consideration. Physical layer security has been recognized as a complementary promising technology to the traditional cryptographic approach to alleviate the eavesdropping problem\cite{Xu2019ExploitingPS,Wang2020JointOO,Zhou2020OffloadingOF,He2020Physical,Wu2020Resource,Xu2021Joint,Liu2021Physical}.
For example, the authors in \cite{Xu2019ExploitingPS} investigated the secure computation offloading in a multiuser multicarrier system.
Following this research, this topic has been extensively studied by applying new technologies for further secrecy performance enhancement, such as multiple antennas\cite{Wang2020JointOO}, cooperative jamming \cite{Zhou2020OffloadingOF}, full-duplex \cite{He2020Physical}, and nonorthogonal multiple access (NOMA) \cite{Wu2020Resource}.
Additionally, the secure MEC technology is extended in different network scenarios, including unmanned aerial vehicle (UAV) \cite{Xu2021Joint} and vehicle networks \cite{Liu2021Physical}.
To the best of the authors' knowledge, the secure transmission in a multi-access MEC network has not been touched yet. This motivates the research of this paper.

In particular, in this paper, we consider a multi-access MEC network where multiple users offload computation tasks to multiple BSs with MEC servers integrated, in the existence of a malicious eavesdropper.
Specifically, we exploit physical layer security technology to improve the security of the system. Each user can divide its task into several parts, one of which is computed locally, and the others are securely offloaded to different MEC servers. Under this setup, we aim to maximize the secrecy offloading rate of users by optimizing the communication and computation resource allocation, subject to the newly introduced subcarrier allocation constraints.
Although the formulated problem is non-convex and difficult to be solved, we propose an efficient algorithm employing the Lagrange dual method and iterative update approach.
Finally, we validate the superior performance of our proposed design over other benchmark schemes via numerical results.

\vspace{-0.3cm}
\section{System Model and Problem Formulation}
In this paper, we consider an orthogonal frequency division multiple access (OFDMA)-based MEC system with $N$ subcarriers.
There are $K$ users, $M$ MEC servers belonged to different BSs, and one eavesdropper denoted as $E$, each of which equips with one antenna. Denote $\mathcal{K}=\{1,\cdots,K\}$, $\mathcal{M}=\{1,\cdots,M\}$ and $\mathcal{N}=\{1,\cdots,N\}$ as the sets of users, MEC servers, and subcarriers with the bandwidth $B$, respectively. A tuple of three parameters $\{s_k,c_k,T_k\}, k\in\mathcal{K}$ is employed to describe the task of user $k$, where $s_k$ indicates the amount of input data to be processed, $c_k$ represents the number of CPU cycles for computing 1-bit of input data, $T_k$ is the maximum tolerable latency.
Due to the limited computation and energy resource at user $k$, it can access to multiple MEC servers for potential load balance \cite{Lin2019Joint}, and offload $\lambda_k^m\in[0,1], m\in\mathcal{M}, k\in\mathcal{K}$ portion of data to MEC server $m$, while the residual $(1-\sum_{m\in \mathcal{M}}\lambda_k^m)s_k$ is processed locally. Let $f_k^l$ and $f_k^m, k\in\mathcal{K}, m\in\mathcal{M}$ (in CPU cycle number per second) denote the allocated CPU frequencies by user $k$ and MEC server $m$, respectively. Since the computation capability of the MEC server is limited, the CPU frequencies allocated to its associated users should not exceed the computation capacity, denoted by $F_m$, i.e., $\sum_{k\in \mathcal{K}}f_k^m\leq F_m, \forall m\in\mathcal{M}$.

\subsection{Latency}\label{Latency}
\subsubsection{Local Computing Delay} To process the residual input bits $(1-\sum_{m\in M}\lambda_k^m)s_k$ at user $k\in\mathcal{K}$, the corresponding local computing time is $t_k^l=c_k\left(1-\sum_{m\in \mathcal{M}}\lambda_k^m\right)s_k/f_k^l$.
\subsubsection{Secure Communication and Processing Delays} The subcarriers are assigned to different users in the OFDMA system to avoid the intra-cell interference.
Denote $\{x_k^{n,m}\}$ as the indicator of subcarrier allocation.
To be specific, $x_k^{n,m}=1$ represents that subcarrier $n$ is assigned to user $k$ for offloading its task to MEC server $m$. Otherwise, $x_k^{n,m}=0$. Let $\{\hat{g}_k^n\}$ and $\{h_k^{n,m}\}$ denote the channels from user $k\in\mathcal{K}$ to the eavesdropper and MEC server $m\in\mathcal{M}$ via subcarrier $n\in\mathcal{N}$, respectively. Denote $\tilde{g}_k^n\triangleq\frac{\hat{g}_k^n}{\sigma^2}$ and $\tilde{h}_k^{n,m}\triangleq\frac{h_k^n}{\sigma^2}$ as the corresponding channel-power-to-noise ratios, where $\sigma^2$ is the variance for additive white Gaussian noise. Assume that the MEC server perfectly knows the channel state information (CSI) of $\tilde{h}_k^{n,m}$, but partially knows that of $\tilde{g}_k^n$ \cite{Xu2019ExploitingPS}. As commonly adopted in the physical layer security literature, we consider the deterministic CSI uncertainty model for $\tilde{g}_k^n$, where $\tilde{g}_k^n=\bar{g}_k^n+\Delta{g}_k^n$, $k\in\mathcal{K}$, $n\in\mathcal{N}$. Here, $\bar{g}_k^n$ denotes the estimation of $\tilde{g}_k^n$ at the MEC server, and $\Delta{g}_k^n$ denotes the estimation error, i.e., $|\Delta{g}_k^n|\leq\epsilon$, bounded by a possible value $\epsilon\geq 0$ (also known by the MEC server). Thus, the secrecy offloading rate (in bits/sec) from user $k$ to MEC server $m$ is given as
\begin{equation}\label{secrecy rate}
r_k^m=\sum\nolimits_{n\in \mathcal{N}}x_k^{n,m}r_k^{n,m},
\end{equation}
where $r_k^{n,m}\triangleq B\left[\log_2\left(1+p_k^n \tilde{h}_k^{n,m}\right)\!-\!\log_2\left(1+p_k^n \tilde{g}_k^n\right)\right]^+$, $[x]^+ \triangleq \max[x,0]$, and $p_k^n$ is the transmit power of user $k\in\mathcal{K}$ over subcarrier $n\in\mathcal{N}$.
Accordingly, the offloading time $t_{k,m}^\mathrm{off}$ for user $k$ is given by $ t_{k,m}^\mathrm{off}=t_{k,m}^s+t_{k,m}^c$, which mainly consists of two parts: the secure offloading time $t_{k,m}^s\triangleq\frac{s_k\lambda_k^m}{r_k^m}$ and the computing time $t_{k,m}^c\triangleq\frac{c_m s_k\lambda_k^m}{f_k^m}$ at MEC server $m$.

Due to the parallel computing at the users and MEC servers, the total latency for user $k$ is $\max\left\{t_k^l,\max_{m\in\mathcal{M}}\left\{t_{k,m}^\mathrm{off}\right\}\right\}$, which depends on the largest one among $t_k^l$ and $t_{k,m}^\mathrm{off}$.\footnote{In practice, the MEC-integrated BS is usually powered by grid with almost unlimited energy, which can provide a sufficiently large transmit power. Furthermore, the amount of output data from MEC server to user is usually much less than that of the input data. Hence, the time consumed for delivering the computed results is negligible \cite{Zhao2021TVT}.}

\vspace{-0.3cm}
\subsection{Energy Consumption}
Since MEC servers are often powered by grid electricity, we only considered the energy consumption of the users.

\subsubsection{Local Computing Mode} The computation energy efficiency coefficient related to the processor's chip of user $k\in\mathcal{K}$ is characterized by $\eta_k$, then the power consumption of the processor is modeled as $\eta_k {f_k^l}^3$ (in joule per second)\cite{Wang2016MobileEdgeCP,Zhao2021TVT}. Thus, the energy consumption of user $k\in\mathcal{K}$ for local computing is $ E_k^l=\eta_k {f_k^l}^3 t_k^l=\eta_k c_k\left(1-\sum_{m\in \mathcal{M}}\lambda_k^m\right) s_k {f_k^l}^2$.

\subsubsection{Computation Offloading Mode} For the data offloaded to MEC servers, the energy consumption of user $k$ for transmission is $ E_k^\mathrm{off}=\sum_{m\in \mathcal{M}}t_{k,m}^s\sum_{n\in \mathcal{N}}x_k^{n,m}\tilde{p}_k=\sum_{m\in \mathcal{M}}\frac{s_k\lambda_k^m\sum_{n\in \mathcal{N}}x_k^{n,m}\tilde{p}_k}{\sum\nolimits_{n\in \mathcal{N}}x_k^{n,m}r_k^{n,m}}$,
where $\tilde{p}_k=p_k^n+\bar{p}_k$, and $\bar{p}_k$ is the constant average circuit power consumption to transmit signal processing at user $k\in\mathcal{K}$ \cite{Xu2021TWC}.
Therefore, the total energy consumption of user $k$ is given by $E_k=E_k^l+E_k^\mathrm{off}$.

\vspace{-0.3cm}
\subsection{Problem Formulation}
Define $\boldsymbol{Q}\triangleq\left\{p_k^n\right\}\in \mathbb{R}^{K\times N}$, $\boldsymbol{X}\triangleq\left\{x_k^{n,m}\right\}\in \mathbb{R}^{K\times NM}$, $\boldsymbol{\Lambda}\!\triangleq\!\{\lambda_k^m\}\in \mathbb{R}^{K\times M}$, $\boldsymbol{f}\triangleq\{f_k^l\}\in \mathbb{R}^{K}$ and $\boldsymbol{F}\triangleq\left\{f_k^m\right\}\in \mathbb{R}^{K\times M}$ for notation convenience.
We aim to maximize the secrecy offloading rate of users via jointly optimizing communication (transmit power $\boldsymbol{Q}$, offloading ratio $\boldsymbol{\Lambda}$ and subcarriers $\boldsymbol{X}$), and computation resource ($\boldsymbol{f}$, $\boldsymbol{F}$), which is formulated as
\begin{subequations}\label{OP1}
 \begin{align}
 \mathbf{P_0}:&\max_{\left\{\boldsymbol{Q},\boldsymbol{X},\boldsymbol{\Lambda},\boldsymbol{f},\boldsymbol{F}\right\}}~\sum_{k\in\mathcal{K}}\sum_{n\in\mathcal{N}}\sum_{m\in\mathcal{M}}x_k^{n,m}r_k^{n,m}\\
 ~ \mathrm{s.t.} ~& \max\left\{t_k^l,\max\nolimits_{m\in\mathcal{M}}\left\{t_{k,m}^\mathrm{off}\right\}\right\}\leq T_k,~\forall k\in\mathcal{K},\label{OP1-C1}\\
 & E_k^l+E_k^\mathrm{off}\leq E_k,~\forall k\in\mathcal{K},\label{OP1-C2}\\
 & |\Delta{g}_k^n|\leq\epsilon,~\forall k\in\mathcal{K},n\in\mathcal{N},\label{OP-C2}\\
 & 0\leq\sum\nolimits_{m\in\mathcal{M}}\lambda_k^m\leq 1,\lambda_k^m \in [0,1],~\forall k\in\mathcal{K},\label{OP1-C4}\\
 & 0\leq p_k^n,~\forall {k\in\mathcal{K},n\in\mathcal{N}},\label{OP1-C5}\\
 & 0\leq\sum\nolimits_{n\in \mathcal{N}}\sum\nolimits_{m\in \mathcal{M}}x_k^{n,m}p_k^n\leq p_k^\mathrm{max},~\forall k\in\mathcal{K},\label{OP1-C6}\\
 & x_k^{n,m}\in \{0,1\},~\forall{k\in\mathcal{K},n\in\mathcal{N},m\in\mathcal{M}},\label{OP1-C7}\\
 & \sum\nolimits_{m\in\mathcal{M}}\sum\nolimits_{K\in\mathcal{K}}x_k^{n,m}\leq 1,~\forall n\in\mathcal{N},\label{OP1-C8}\\
 & 0\leq f_k^m,~\forall {k\in\mathcal{K},\forall m\in\mathcal{M}},\label{OP1-C10}\\
 & \sum\nolimits_{k\in\mathcal{K}}f_k^m\leq F_m,~\forall m\in\mathcal{M},\label{OP1-C11}\\
 & 0\leq f_k^l\leq F_k,~\forall k\in\mathcal{K},\label{OP1-C12}
 \end{align}
\end{subequations}where $E_k$ is the maximum energy budget and $p_k^\text{max}$ is the maximum transmit power at user $k$.
Constraint \eqref{OP1-C1} requests to complete the tasks under the latency requirements of users. Constraints \eqref{OP1-C2} and \eqref{OP1-C6} ensure that the total energy consumption and power allocation among subcarriers cannot exceed the maximum energy budget and transmit power of each user. Each subcarrier can only be used by one User-MEC pair to avoid the same-frequency interference in \eqref{OP1-C8}, and the sum of allocated CPU frequencies must be less than the computation capability of MEC server in \eqref{OP1-C11}.

\begin{figure*}
\begin{align}
\mathcal{L}(\boldsymbol{W}, \boldsymbol{\alpha},\boldsymbol{\beta}, \boldsymbol{\gamma}, \boldsymbol{\theta}, \boldsymbol{\mu}, \boldsymbol{\psi}, &\boldsymbol{\varphi})=\sum\nolimits_{k\in \mathcal{K}}\sum\nolimits_{n\in \mathcal{N}}\sum\nolimits_{m\in \mathcal{M}}x_k^{n,m}\bigg[\left(1\!+\!\psi_k^m\right)r_k^{n,m}\!-\!{\gamma_k s_k \lambda_k^m \left(p_k^n+\bar{p}_k\right)}/{\phi_k^m}\!-\!\theta_k p_k^n\bigg]\nonumber\\
&-\sum\nolimits_{k\in \mathcal{K}}\sum\nolimits_{m\in \mathcal{M}}\beta_k^m\left(s_k\lambda_k^m/\phi_k^m\!+\!{c_m s_k\lambda_k^m/{f_k^m}}\!-\!T_k\right)\!+\!\mu_m f_k^m\!+\!\psi_k^m\phi_k^m\!+\!\sum\nolimits_{m\in \mathcal{M}}\mu_m F_m\label{OP3-5}\\
&-\sum\nolimits_{k\in \mathcal{K}}\bigg[\alpha_k\left(t_k^l\!-\!T_k\right)\!+\!\gamma_k\left(E_k^l\!-\!E_k\right)\!-\!\theta_k p_k^{\mathrm{max}}\!+\!\varphi_k\left(f_k^l\!-\!F_k\right)\vphantom{\sum_{n\in \mathcal{N}}}\bigg].\nonumber\\
\hline
\nonumber
\end{align}\vspace{-1.5cm}
\end{figure*}

\section{Resource-Allocation Based Algorithm}
Due to the coupled variants in both the constraints and the objective function, it is intractable to deal with $\mathbf{P_0}$ in its current form. To tackle this issue, we first consider the worst-case secrecy offloading rate, i.e., $g_k^n=\bar{g}_k^n+\epsilon,~\forall {k\in\mathcal{K},n\in\mathcal{N}}$, which is given as
\begin{equation}
\!\tilde{r}_k^m\!=\!\min\nolimits_{|\Delta{g}_k^n|\leq\epsilon}\!\sum\nolimits_{n\in \mathcal{N}}x_k^{n,m}{r_k^{n,m}}\!=\!\!\sum\nolimits_{n\in \mathcal{N}}x_k^{n,m}\tilde{r}_k^{n,m},\!
\end{equation}
where $\tilde{r}_k^{n,m}\triangleq B\left[\log_2\left(1+p_k^n \tilde{h}_k^{n,m}\right)-\log_2\left(1+p_k^n g_k^n\right)\right]^+$.

Therefore, we propose to maximize the explicit and mathematically tractable function $\tilde{r}_k^m$ to obtain a lower bound for the secrecy offloading rate.
Furthermore, according to \cite{ZhongSecure2018}, we can equivalently remove the $[\cdot]^+$ operator in the objective function and approximate problem $\mathbf{P_0}$ as the following problem:
\begin{subequations}\label{OP1-1}
\begin{align}
 &\mathbf{P_0'}:\max_{\left\{\boldsymbol{Q},\boldsymbol{X},\boldsymbol{\Lambda},\boldsymbol{f},\boldsymbol{F}\right\}}~\sum_{k\in\mathcal{K}}\sum_{n\in\mathcal{N}}\sum_{m\in\mathcal{M}}x_k^{n,m}\bar{r}_k^{n,m}\\
 \mathrm{s.t.} ~& \max\left\{t_k^l,\max\nolimits_{m\in\mathcal{M}}\left\{t_{k,m}^\mathrm{off}(\bar{r}_k^m)\right\}\right\}\leq T_k,~\forall k\in\mathcal{K},\label{OP1-1-C1}\\
 &\eqref{OP1-C2}-\eqref{OP1-C12},\nonumber
 \end{align}
 \end{subequations}
where $\bar{r}_k^{n,m}\triangleq B\left[\log_2\left(1+p_k^n \tilde{h}_k^{n,m}\right)\!-\!\log_2\left(1+p_k^n g_k^n\right)\right]$.

Next, we transform the formulated problem $\mathbf{P_0'}$ into five subproblems: 1) $\mathbf{P_1}$, offloading ratio optimization; 2) $\mathbf{P_2}$, communication resource allocation; 3) $\mathbf{P_3}$, MEC servers computation resource optimization; 4) $\mathbf{P_4}$, user computation resource optimization; 5) $\mathbf{P_5}$, auxiliary variable update. Then, we optimize them in an iterative manner while keeping the other variables fixed.

\vspace{-0.3cm}
\subsection{Offloading Ratio Optimization}\label{off}
Since the objective function of $ \mathbf{P_0'}$ is independent with $\boldsymbol{\Lambda}$, the corresponding problem can be treated as a feasible problem with given $\left(\boldsymbol{X},\boldsymbol{Q},\boldsymbol{F},\boldsymbol{f}\right)$, and the feasible solution of offloading ratio can be derived from
 \begin{align}\label{P1}
\mathbf{P_1}:&~\text{find}~ \lambda_k^m,\\
\mathrm{s.t.} ~& \eqref{OP1-1-C1},\eqref{OP1-C2},\eqref{OP1-C4},\nonumber
 \end{align}
which is a linear programming problem with respect to $\lambda_k^m,~\forall k\in\mathcal{K},m\in\mathcal{M}$, and can be solved by optimization tools such as CVX.

With the obtained feasible solution $\boldsymbol{\Lambda}^\star$, $\mathbf{P_0'}$ is still mixed-integer nonlinear programming, which is generally prohibitively to find the optimal solution due to the high computation complexity. Fortunately, the duality gap becomes zero in multi-carrier systems as the number of subcarriers goes to infinity according to the time-sharing condition. Thus, the asymptotically optimal solution for a non-convex resource allocation problem can be achieved in the dual domain when the number of subcarriers becomes large \cite{Zhao2021TVT,Wei2006Dual}.

However, $\mathbf{P_0'}$ cannot be transformed into the dual domain directly since $\sum_{n\in \mathcal{N}}x_k^{n,m}\bar{r}_k^{n,m}$ is in the denominator of constraint \eqref{OP1-1-C1}. Therefore, a new non-negative auxiliary variable $\boldsymbol{\Phi}\triangleq\{\phi_k^m\},\forall k\in\mathcal{K},m\in\mathcal{M}$ will be introduced to transform $\mathbf{P_0'}$ into the following problem
\begin{subequations}\label{OP2}
 \begin{align}
\mathbf{P_0''}:&\max_{\left\{\boldsymbol{X},\boldsymbol{Q},\boldsymbol{F},\boldsymbol{f},\boldsymbol{\Phi}\right\}}\sum_{k\in\mathcal{K}}\sum_{n\in\mathcal{N}} \sum_{m\in\mathcal{M}}x_k^{n,m}\bar{r}_k^{n,m}\\
\mathrm{s.t.} &~ t_k^l\leq T_k,\forall k\in\mathcal{K},\label{OP2-C0}\\
 &s_k\lambda_k^m/\phi_k^m+{c_m s_k\lambda_k^m}/{f_k^m} \leq T_k,\forall k\in\mathcal{K},m\in\mathcal{M},\label{OP2-C1}\\
 &E_k^l+\!\!\sum_{m\in\mathcal{M}}\frac{s_k \lambda_k^m}{\phi_k^m}\sum_{n\in\mathcal{N}}x_k^{n,m}\tilde{p}_k\leq E_k,\forall k\in\mathcal{K},\label{OP2-C2}\\
 & 0<\phi_k^m\leq\sum\nolimits_{n\in\mathcal{N}}x_k^{n,m}\bar{r}_k^{n,m},\forall {k\in\mathcal{K},m\in\mathcal{M}},\label{OP2-C3}\\
 &\eqref{OP1-C5}-\eqref{OP1-C12},\nonumber
 \end{align}
\end{subequations}
and the Lagrangian function of $\mathbf{P_0''}$ is given at the top of this page, where $\boldsymbol{\alpha}, \boldsymbol{\beta}, \boldsymbol{\gamma}, \boldsymbol{\theta}, \boldsymbol{\mu}, \boldsymbol{\psi}$ and $    \boldsymbol{\varphi}$ are the non-negative Lagrange multipliers for the corresponding constraints.

Define $\mathcal{P}$ as all sets of possible $\boldsymbol{Q}$ that satisfy constraint \eqref{OP1-C5}, $\mathcal{X}$ as all sets of possible $\boldsymbol{X}$ that satisfy constraint \eqref{OP1-C7}, $\mathcal{F}$ as all sets of possible $\boldsymbol{F}$ that satisfy constraint \eqref{OP1-C10}, $\mathcal{Q}$ as all sets of possible $\boldsymbol{f}$ that satisfy constraint $f_k^l>0$, and $\mathcal{S}$ as all sets of possible $\boldsymbol{\phi}$ that satisfy constraint $\phi_k^m>0,~\forall k\in\mathcal{K},m\in\mathcal{M}$. The Lagrange dual function is then defined as
\begin{align}
 & g(\boldsymbol{\alpha}, \boldsymbol{\beta}, \boldsymbol{\gamma}, \boldsymbol{\theta}, \boldsymbol{\mu}, \boldsymbol{\psi}, \boldsymbol{\varphi})=\nonumber\\
 & \max_{\substack{\left\{\boldsymbol{X}\in \mathcal{X},\boldsymbol{Q}\in \mathcal{P},\boldsymbol{F}\in \mathcal{F},\right.\\
 \left.\boldsymbol{f}\in \mathcal{Q},\boldsymbol{\Phi}\in \mathcal{S}\right\}}}
 \mathcal{L}(\boldsymbol{W}, \boldsymbol{\alpha}, \boldsymbol{\beta}, \boldsymbol{\gamma}, \boldsymbol{\theta},\boldsymbol{\mu}, \boldsymbol{\psi}, \boldsymbol{\varphi}),\label{OP2-C4}
\end{align}
where $\boldsymbol{W}\triangleq\{\boldsymbol{X}, \boldsymbol{Q}, \boldsymbol{F}, \boldsymbol{f}, \boldsymbol{\Phi}\}$. Moreover, the Lagrange dual problem is then given by $\max~g(\boldsymbol{\alpha}, \boldsymbol{\beta}, \boldsymbol{\gamma}, \boldsymbol{\theta}, \boldsymbol{\mu}, \boldsymbol{\psi}, \boldsymbol{\varphi})$, where $\boldsymbol{\alpha}, \boldsymbol{\beta}, \boldsymbol{\gamma}, \boldsymbol{\theta}, \boldsymbol{\mu}, \boldsymbol{\psi}, \boldsymbol{\varphi}\succeq\boldsymbol{0}$.

The primal-dual method is employed to obtain the above resource-related variables for the given auxiliary variable and Lagrange multipliers at first, then the auxiliary variable is updated based on its closed form. At last, the corresponding Lagrange multipliers are updated via the subgradient method.

In the following, the other four subproblems are given by
\begin{align}
 \!\!\!\!&\mathbf{P_2}: g(\boldsymbol{\gamma},\boldsymbol{\theta}, \boldsymbol{\psi})=\max_{\boldsymbol{X}\in\mathcal{X},\boldsymbol{Q}\in\mathcal{P}}\!\sum_{k\in \mathcal{K}}\sum_{n\in \mathcal{N}}\sum_{m\in\mathcal{M}}x_k^{n,m}\mathcal{L}_k^{n,m},\\
 \!\!\!\!&\mathbf{P_3}:g(\boldsymbol{\beta},\boldsymbol{\mu})=\max_{\boldsymbol{F}\in\mathcal{F}}\sum\nolimits_{m\in\mathcal{M}}\sum\nolimits_{k\in \mathcal{K}}\mathcal{L}_k^m(f_k^m),\\
 \!\!\!\!&\mathbf{P_4}:g(\boldsymbol{\alpha},\boldsymbol{\gamma},\boldsymbol{\varphi})=
 \max_{\boldsymbol{f}\in\mathcal{Q}}\sum\nolimits_{k\in\mathcal{K}}\mathcal{L}_k(f_k^l),\\
 \!\!\!\!&\mathbf{P_5}:g(\boldsymbol{\beta},\boldsymbol{\gamma},\boldsymbol{\psi})=
 \max_{\boldsymbol{\Phi}\in\mathcal{S}}\sum\nolimits_{k\in\mathcal{K}}\sum\nolimits_{m\in\mathcal{M}}\mathcal{L}_k^m\left(\phi_k^m\right),
\end{align}
where
\begin{align}
 &\mathcal{L}_k^{n,m}\!=\!(1\!+\!\psi_k^m)\bar{r}_k^{n,m}\!-\!(\gamma_k s_k
 \lambda_k^m\!+\!\theta_k \phi_k^m)p_k^n/\phi_k^m,\nonumber\\
 &\mathcal{L}_k^m(f_k^m)\!=\! -\beta_k^m s_k
 \lambda_k^m c_m/{f_k^m}\!-\!\mu_m f_k^m,\nonumber\\
 &\mathcal{L}_k(f_k^l)\!=\!-c_k s_k\left(\!\alpha_k\!+\!\gamma_k\eta_k {f_k^l}^3\!\right)\left(\!1\!-\!\sum\nolimits_{m\in\mathcal{M}}\!\!\lambda_k^m\!\right)/{f_k^l}\!-\!\varphi_k f_k^l,\nonumber\\
 &\mathcal{L}_k^m\left(\phi_k^m\right)\!=\!-\psi_k^m\phi_k^m\!-\!s_k\lambda_k^m\left(\beta_k^m\!+\!\gamma_k\sum\nolimits_{n\in \mathcal{N}}x_k^{n,m}\tilde{p}_k\right)/\phi_k^m,\nonumber
\end{align}
and we can solve them as follows.
\subsection{Communication Resource Allocation}\label{Allocation}

\subsubsection{Subcarriers Allocation}
Given $\{\boldsymbol{Q}, \boldsymbol{F}, \boldsymbol{f}, \boldsymbol{\Lambda}^\star, \boldsymbol{\Phi}\}$, we assume that all subcarriers are assigned to users, i.e., $\sum_{k\in \mathcal{K}}\sum_{m\in \mathcal{M}}x_k^{n,m}\!=\!1$, thus there exists $\{k^\star, m^\star\}$ such that
\begin{equation}\label{OP2-x1}
x_k^{n,m,\star}=\left\{\begin{array}{l}{1,~\text {if}~\{k,m\}=\{k^\star,m^\star\}}\\
{0,~\text{otherwise}}\end{array}\right.,
\end{equation}
to maximize $\mathcal{L}_k^{n,m}$ via $\{k^\star,m^\star\}=\arg\max\nolimits_{k\in\mathcal{K},m\in\mathcal{M}}\mathcal{L}_k^{n,m}$.

\subsubsection{Transmit Power Optimization}
With the newly derived $\boldsymbol{X}^\star$ and employing the Karush-kuhn-Tucker (KKT) conditions, the condition $\frac{\partial\mathcal{L}_k^{n,m}\left({p}_k^n\right)}{\partial {p}_k^n}=0$ is both necessary and sufficient for transmit power assignment optimality, which can be obtained as follows.
\begin{theorem}
The optimal transmit power is
\begin{align}\label{OP2-p1}
 ~p_k^{n,\star}=&\Bigg\{\Bigg[\frac{4\tilde{h}_k^{n,m} g_k^n\left(\tilde{h}_k^{n,m}-g_k^n\right)\left(1+\psi_k^m\right)\phi_k^m B}{\left(\gamma_k s_k \lambda_k^m+\theta_k\phi_k^m\right)\ln2}\nonumber\\
 &\!+\!\left(\tilde{h}_k^{n,m}\!-\!g_k^n\right)^2\vphantom{\frac{4h_k^{\frac{1}{2}}g\left(h_k-\tilde{g}^m\right)(1+\psi_k^m)\phi_k^m B}{\left(\gamma_k s_k \lambda_k^m+\theta_k\phi_k^m\right)\ln2}}\Bigg]^{\frac{1}{2}}\!\!-\!\left(\tilde{h}_k^{n,m}\!+\!g_k^n\right)\!\Bigg\}/{2\tilde{h}_k^{n,m}g_k^n}.
 \end{align}
\end{theorem}
\begin{IEEEproof}
It can be obtained from the formula of roots for a quadric equation, which is deduced from $\frac{\partial\mathcal{L}_k^{n,m}\left({p}_k^n\right)}{\partial {p}_k^n}=0$.
\end{IEEEproof}
\begin{remark}
To maximize the secrecy offloading rate of users, user $k\in \mathcal{K}$ will distribute more power on subcarrier $n \in \mathcal{N}$ when the eavesdropper has a worse channel gain $g_k^n$, and the assigned power is proportional to the subcarrier bandwidth $B$.
\end{remark}

\vspace{-0.3cm}
\subsection{MEC Servers Computation Resource Optimization}\label{MEC resource}
With the newly obtained $\left(\boldsymbol{X}^\star,\boldsymbol{P^\star},\boldsymbol{\Lambda}^\star\right)$ and given $(\boldsymbol{f},\boldsymbol{\phi})$, the optimal computation resource assigned to user $k$ can be derived according to the KKT conditions and solving
\begin{equation}
 \frac{\partial\mathcal{L}_k^m\left(f_k^m\right)}{\partial f_k^m}=\frac{\beta_k^m s_k\lambda_k^m c_m}{(f_k^m)^2}-\mu_m=0.\label{OP3-C1}
\end{equation}
Correspondingly, the optimal solution is
\begin{equation}\label{f_k^m}
     {f_k^m}^\star=\sqrt{{\beta_k^m s_k\lambda_k^m c_m}/{\mu_m}}.
\end{equation}
\begin{remark}
It can be observed when user $k \in \mathcal{K}$ decides to offload more data to MEC $m \in \mathcal{M}$, i.e., a larger $\lambda_k^m$, MEC $m$ will assign more computation capability $f_k^m$ to deal with the offloaded task to satisfy its latency requirement.
\end{remark}

\vspace{-0.3cm}
\subsection{User Computation Resource Optimization}\label{user resource}
With the newly obtained $\left(\boldsymbol{X}^\star,\boldsymbol{P^\star},\boldsymbol{\Lambda}^\star,\boldsymbol{F^\star}\right)$, the optimal capability assignment for users $\boldsymbol{f^\star}$ can be obtained by
\begin{align}
 \frac{\partial\mathcal{L}_k(f_k^l)}{\partial f_k^l}&= \frac{\alpha_k c_k s_k(1-\sum\nolimits_{m\in\mathcal{M}}\lambda_k^m)}{{f_k^l}^2}\nonumber\\-&2\gamma_k\eta_k\big(1-
 \sum\nolimits_{m\in\mathcal{M}}\lambda_k^m\big)s_k c_k f_k^l-\varphi_k=0,\label{OP3-C2}
\end{align}
according to KKT conditions with given $\boldsymbol{\phi}$. Furthermore, a cubic equation of $f_k^l$ can be deduced from Eq. \eqref{OP3-C2}, and solved via the secant algorithm.

\vspace{-0.3cm}
\subsection{Auxiliary Variable Update}\label{SCM}
With the newly derived $\left(\boldsymbol{X}^\star,\boldsymbol{P^\star},\boldsymbol{\Lambda}^\star,\boldsymbol{F^\star},\boldsymbol{f}^\star\right)$, we can solve
\begin{equation}
 \frac{\partial\mathcal{L}_k^m\left(\phi_k^m\right)}{\partial\phi_k^m}=\frac{\beta_k^m s_k\lambda_k^m+s_k\lambda_k^m\gamma_k\sum\nolimits_{n\in \mathcal{N}}x_k^{n,m}\tilde{p}_k}{{\phi_k^m}^2}-\psi_k^m=0,\label{OP3-C4}
\end{equation}
to get the optimal solution of ${\phi_k^m}^\star$, given by
\begin{equation}\label{OP2-phi1}
 {\phi_k^m}^\star\!=\!\sqrt{\bigg(\beta_k^m s_k\lambda_k^m\!+\!s_k\lambda_k^m\gamma_k\sum\nolimits_{n\in \mathcal{N}}x_k^{n,m}\tilde{p}_k\bigg)/{\psi_k^m}}.
\end{equation}
\subsection{Lagrange Multipliers Update}
With the achieved $\boldsymbol{W}^\star$, we
start to update the Lagrange multiplier ($ \boldsymbol{\alpha}, \boldsymbol{\beta}, \boldsymbol{\gamma}, \boldsymbol{\mu}, \boldsymbol{\psi}, \boldsymbol{\varphi}$). Since the Lagrange dual problem is always convex, the subgradient method can be applied to update these variables.

\vspace{-0.3cm}
\subsection{Convergence and Complexity}
Since each subproblem can be solved according to the above procedures, our proposed algorithm can tackle $\mathbf{P_0'}$ with asymptotically optimal solutions \cite{Wei2006Dual,1999Nonlinear}, which has a total computation complexity $\mathcal{O}(K^{3}M^{3}L_{\rm iter})$, where $L_{\rm iter}$ is
the number of alternating optimization iterations \cite{Zhao2021TVT,2020AerialGroundCostzhan}. The details of our proposed algorithm are given in Algorithm \ref{Algorithm1}.

\begin{algorithm}[h]
	\caption{\bf{:} Proposed Algorithm}\label{Algorithm1}
	{\bf Initialization:}
	Initialize $\left(\boldsymbol{W},\boldsymbol{\alpha},\boldsymbol{\beta},\boldsymbol{\gamma},\boldsymbol{\theta},\boldsymbol{\mu},\boldsymbol{\psi},\boldsymbol{\varphi}\right)$ and the algorithm accuracy indicators $\epsilon_1$, where $\epsilon_1$ is a very small constant for controlling accuracy, set loop variable $z=0$, and denote $z_\text{max}$ as the maximum number of iterations.
	\begin{algorithmic}[1]
 \Repeat ~(from 1 to 14)
 \State Find the feasible solution of offloading ratio via solving Problem $\mathbf{P_1}$.
 \Repeat ~(from 3 to 12)
 \Repeat ~(from 4 to 10)
 \State Determine subcarriers allocation via Eq. \eqref{OP2-x1}.
 \State Determine transmit power via Eq. \eqref{OP2-p1}.
 \State Optimize MEC servers computation resource allocation via Eq. \eqref{f_k^m}.
 \State Optimize user computation resource allocation via Eq. \eqref{OP3-C2} by the secant algorithm.
 \State Update $\boldsymbol{\phi}$ according to Eq. \eqref{OP2-phi1}.
 \Until Lagrangian function converges.
 \State Update $\boldsymbol{\alpha}$, $\boldsymbol{\beta}$, $\boldsymbol{\gamma}$, $\boldsymbol{\theta}$, $\boldsymbol{\mu}$, $\boldsymbol{\psi}$ and, $\boldsymbol{\varphi}$.
 \Until {$\boldsymbol{\alpha}$, $\boldsymbol{\beta}$, $\boldsymbol{\gamma}$, $\boldsymbol{\theta}$, $\boldsymbol{\mu}$, $\boldsymbol{\psi}$ and, $\boldsymbol{\varphi}$ converge.}
 \State $z = z + 1$.
 \Until the difference successive values of the objective function is less than $\epsilon_1$ or $z>z_{\rm max}$.
 \end{algorithmic}	
\end{algorithm}
\begin{figure*}[!t]
\begin{minipage}[!h]{0.33\linewidth}
\centering
\includegraphics[width=2.1in]{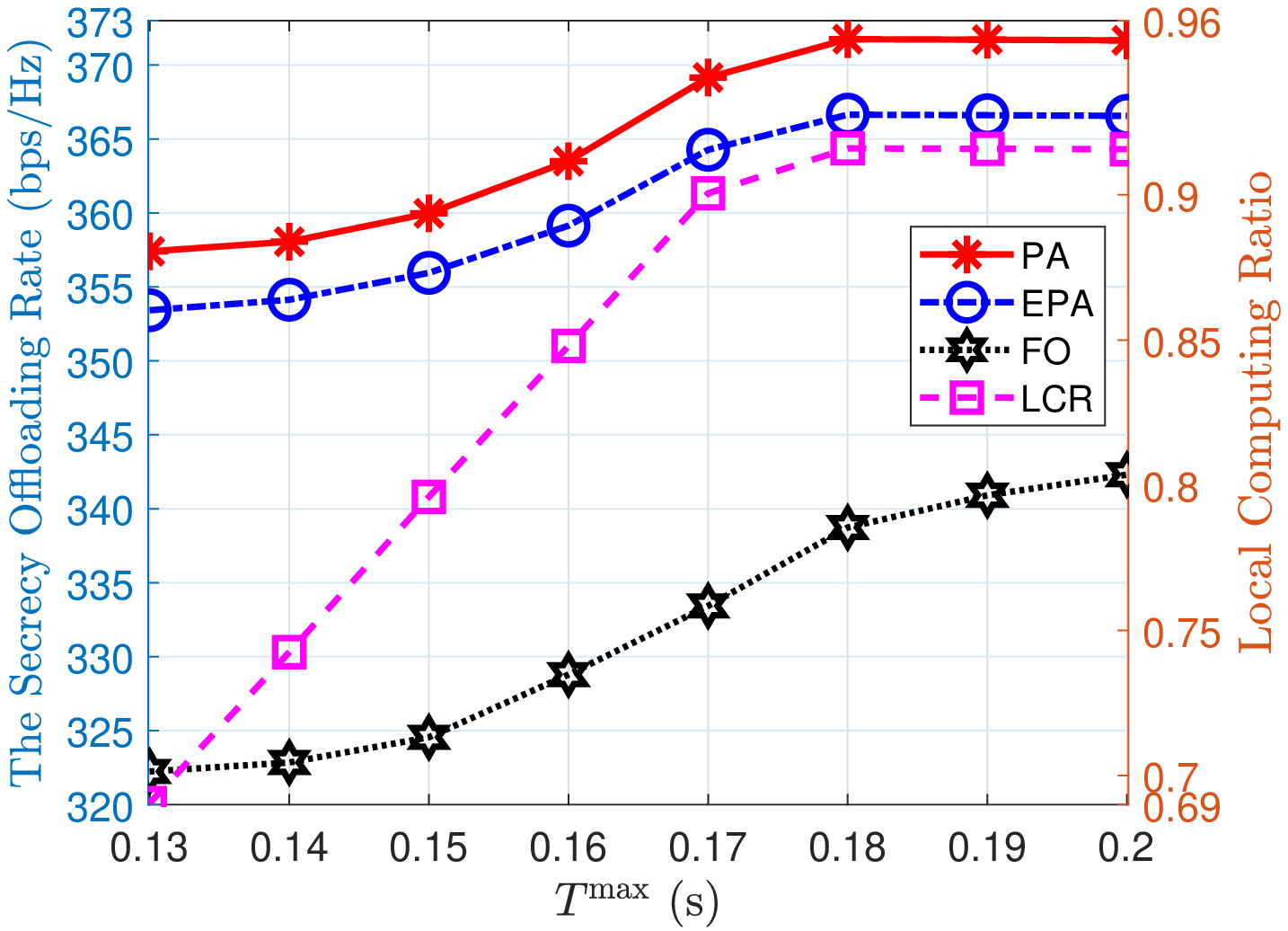}
\caption{Secrecy offloading rate and local computing ratio versus tasks' latency requirements.}
\label{fig:side:a}
\end{minipage}%
\begin{minipage}[!h]{0.33\linewidth}
\centering
\includegraphics[width=2.1 in]{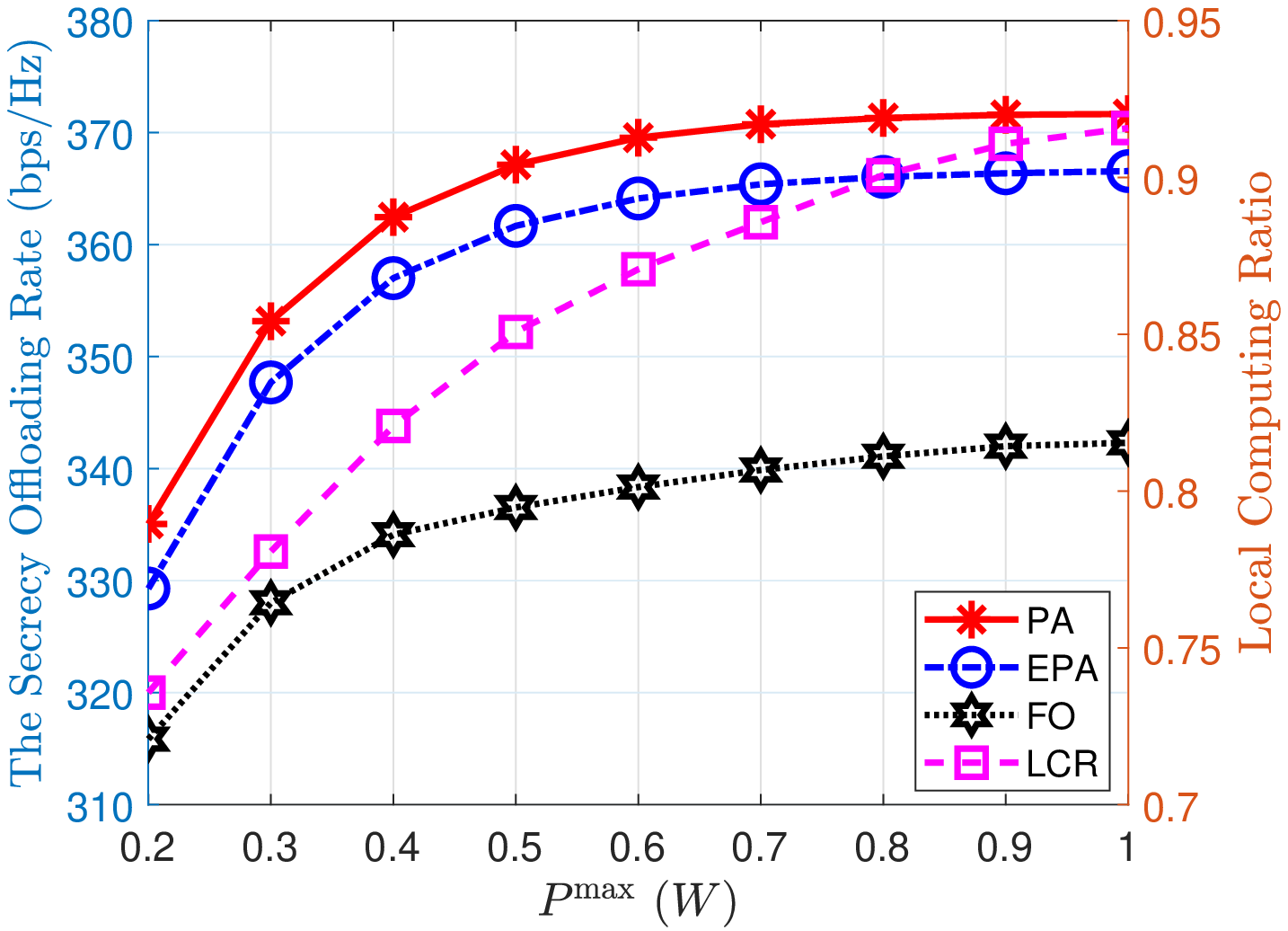}
\caption{Secrecy offloading rate and local computing ratio versus users' maximum transmit power.}
\label{fig:side:b}
\end{minipage}
\begin{minipage}[!h]{0.33\linewidth}
\centering
\includegraphics[width=2.1 in]{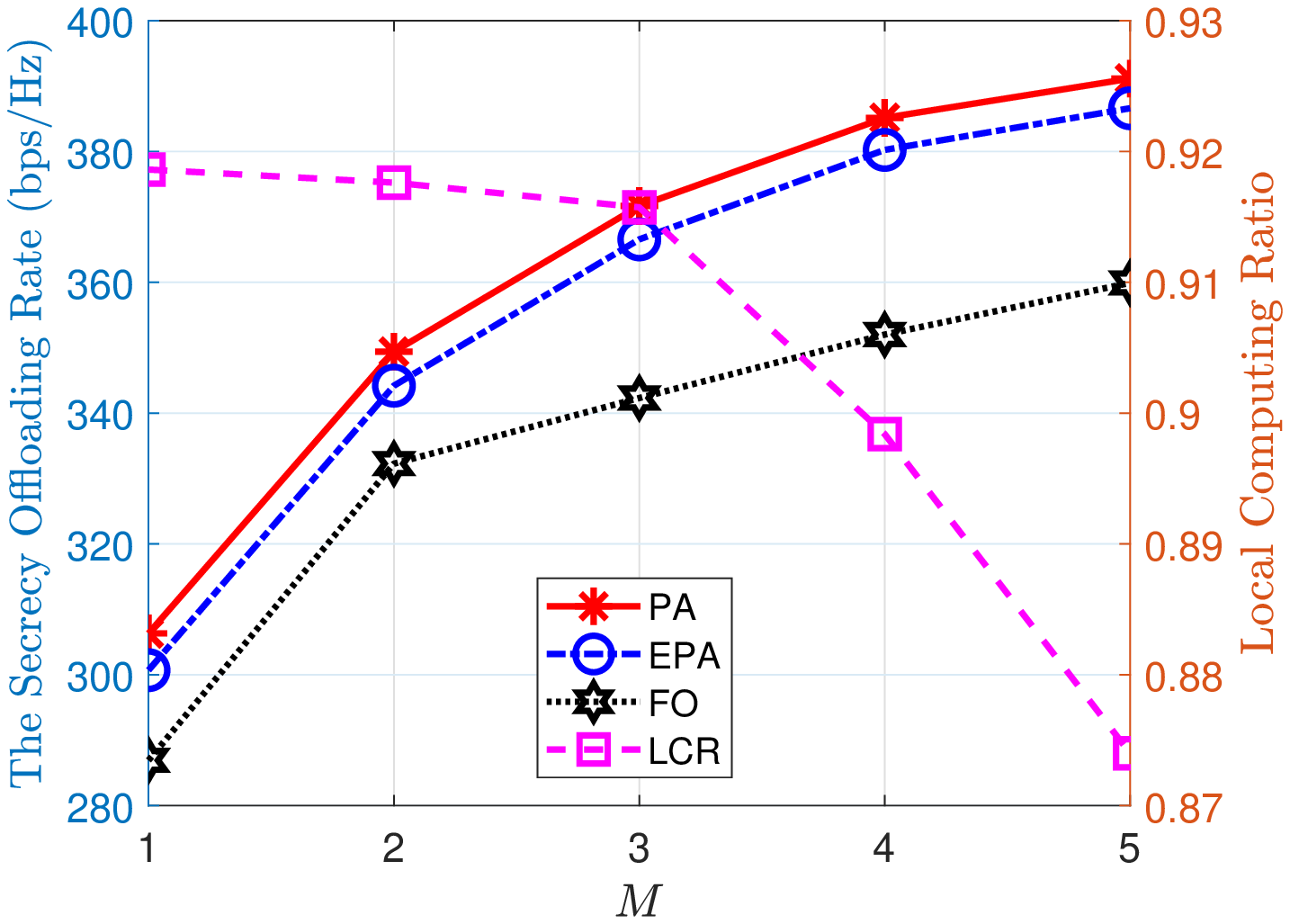}
\caption{Secrecy offloading rate and local computing ratio versus MEC server number.}
\label{fig:side:c}
\end{minipage}\vspace{-0.6cm}
\end{figure*}

\vspace{-0.3cm}
\section{Simulation And Numerical Results}
Numerical results are given to demonstrate
the superior performance of our Proposed Algorithm (\textbf{PA}) compared with the reference schemes: 1) Equal power allocation (\textbf{EPA}), where the user equivalently distributes the transmit power among its allocated subcarriers; 2) Full offloading (\textbf{FO}), where the users fully offload their tasks to MEC servers for processing.

In the simulation, unless otherwise stated, we use the following settings: the channel model follows Rayleigh fading and the average channel power gain is specified by $\beta_0(d/d_0)^{-\alpha}$, where $\beta_0 \!=\!\! -30$~dBm corresponds to the pathloss at a reference distance of $d_0\!\!\!=\!\!\!1$~m, $d$ represents the distance between the respective transmit-receive nodes, and $\alpha\!=\!2.1$ is the pathloss exponent, respectively. Moreover, we set $z_\text{max} \!=\! 100$, $T_k\!=\!T^\text{max}\!=\!0.2$~s, $K\!=\!5$, $N \!=\!64$, $M\!=\!3$, $\sigma^2\!=\!10^{-10}$~mW, $\eta_k \!=\! 10^{-24}$, $c_k \!=\! 1100$~cycles/bit, $F_k\!=\!0.7$~GHz, $F_m\!=\!1.1$~GHz, $\bar{p}_k\!=\!10^{-0.3}$~mW, $p_k^\text{max}\!=\!P^\text{max}\!\!=\!\! 10^3$~mW, $B \!=\! 12.5$~KHz, $d_k\!=\!9\times 10^5$~bits, $d_k^m$ and $d_k^e$ range from $[50,55]$~m, $\forall k\in\mathcal{K}, m\in\mathcal{M}$ \cite{Zhao2021TVT,Xu2021TWC}.

In Fig. \ref{fig:side:a}, we present the secrecy offloading rate and the local computing ratio versus the latency requirements of tasks. It is observed that both the secrecy offloading rate and the local computing ratio (LCR) increase with the gradually decreasing latency requirement of tasks. This is due to the fact that when the latency requirement of tasks becomes looser, the user will have a more sufficient local computing time, leading to a lower local CPU frequency $f_k^l$ and a less local computing energy consumption. Hence, more energy can be used for transmission, which in turn allows a higher transmit power and LCR.
Additionally, since the secrecy offloading rate is proportional to the transmit power, a higher transmit power means a higher secrecy offloading rate.
It is also observed that \textbf{PA} can obtain additional 1.11\%--1.39\% and 15.05\%--17.35\% of the secrecy offloading rate compared to \textbf{EPA} and \textbf{FO} with diverse latency requirements.

In Fig. \ref{fig:side:b}, we investigate the secrecy offloading rate and the local computing ratio versus the maximum transmit power of the users.
It is observed that both the secrecy offloading rate and the LCR increase with the increasing users' maximum transmit power, and get saturated when $P^{\rm max}\geq 0.9$~W. It is due to the fact that a larger maximum transmit power of the users represents a larger transmit power of users. Within the same total energy consumption for transmission, a larger instantaneous transmit power means a higher rate and a less transmission time, but less total transmission data, thus resulting in the increase of the LCR.
It is also observed that \textbf{PA} can derive additional 1.30\%--1.75\% and 6.08\%--9.22\% of the secrecy offloading rate compared with \textbf{EPA} and \textbf{FO} with the increase of the maximum transmit power.

In Fig. \ref{fig:side:c}, we explore the secrecy offloading rate and the local computing ratio versus the number of MEC servers.
It is observed that the secrecy offloading rate increases and the LCR decreases with the increase of the MEC server number.
It is due to the fact that, as the number of MEC servers increases, the users can enjoy better channel gains with a higher probability to offload more data within the same period of time, and thus obtain a higher secrecy rate and lower LCR.

\section{Conclusion}
This letter investigated the secure computation offloading in a multi-access MEC network by leveraging the physical layer security technology. We focused on maximizing the secrecy offloading rate of the users by optimizing the communication and computation resource allocation, while satisfying the newly introduced subcarrier allocation constraints.
How to extend our work to other setups (e.g., multiple antennas and full-duplex) are interesting future directions worth investigating.

\footnotesize
\bibliographystyle{IEEEtran}
\bibliography{main}
\end{document}